\documentclass[fleqn,usenatbib]{mnras}

\usepackage{newtxtext,newtxmath}
\usepackage[T1]{fontenc}

\DeclareRobustCommand{\VAN}[3]{#2}
\let\VANthebibliography\thebibliography
\def\thebibliography{\DeclareRobustCommand{\VAN}[3]{##3}\VANthebibliography}

\usepackage{graphicx}	% Including figure files
\usepackage{amsmath}	% Advanced maths commands
\usepackage{hyperref}

\usepackage{threeparttable}
\usepackage{booktabs,caption}
\newcommand{\hMpc}{{\ifmmode{\,h^{-1}{\rm Mpc}}\else{$h^{-1}$Mpc}\fi}}
\newcommand{\hkpc}{{\ifmmode{\,h^{-1}{\rm kpc}}\else{$h^{-1}$kpc}\fi}}
\newcommand{\hMsun}{{\ifmmode{\,h^{-1}{\rm {M_{\odot}}}}\else{$h^{-1}{\rm{M_{\odot}}}$}\fi}}
\newcommand{\Msun}{\,\rm {M_{\odot}}}
\newcommand{\Mstar}{{\ifmmode{\,M_{*}}\else{$M_{*}$}\fi}}
\newcommand{\Mhalo}{{\ifmmode{\,M_{\rm halo}}\else{$M_{\rm halo}$}\fi}}
\newcommand{\ltsima}{$\; \buildrel < \over \sim \;$}
\newcommand{\gtsima}{$\; \buildrel > \over \sim \;$}
\newcommand{\lsim}{\lower.5ex\hbox{\ltsima}}
\newcommand{\gsim}{\lower.5ex\hbox{\gtsima}}

\newcommand{\simba}[0]{\mbox{\textsc{Simba}}}
\newcommand{\hyenas}[0]{\mbox{\textsc{Hyenas}}}
\newcommand{\moxha}[0]{\mbox{\textsc{MOXHA}}}

\newcommand{\caesar}[0]{{ \textsc{Caesar}}}

\newcommand{\GIZMO}[0]{{ \textsc{Gizmo}}}
\newcommand{\music}[0]{{ \textsc{MUSIC}}}

\title[The HYENAS project]{The HYENAS project: a prediction for the X-ray undetected galaxy groups}

\author[Cui et al.]{
Weiguang Cui,$^{1,2,3}$\thanks{E-mail: weiguang.cui@uam.es; Talento-CM fellow}
Fred Jennings,$^{3}$
Romeel Dave,$^{3,4}$
Arif Babul$^{5,6,7}$
and Ghassem Gozaliasl$^{8,9}$
\\
$^{1}$ Departamento de Física Teórica, M-8, Universidad Autónoma de
Madrid, Cantoblanco 28049, Madrid, Spain\\
$^{2}$ Centro de Investigación Avanzada en Física Fundamental,(CIAFF), Universidad Aut\'{o}noma de Madrid, Cantoblanco, 28049 Madrid, Spain\\
$^{3}$ Institute for Astronomy, University of Edinburgh, Royal Observatory, Blackford Hill, Edinburgh EH9 3HJ, UK.\\
$^{4}$ Department of Physics and Astronomy, University of the Western Cape, Robert Sobukwe Rd, Cape Town 7460, South Africa\\
$^{5}$ Department of Physics and Astronomy, University of Victoria, 3800 Finnerty Road, Victoria, V8P 1A1, BC, Canada\\
$^{6}$ Infosys Visiting Chair Professor, Indian Institute of Science, Bangalore, 560012, India\\
$^{7}$ Leverhulme Visiting Professor, Institute for Astronomy, University of Edinburgh, Blackford Hill, Edinburgh, EH9 3HJ, Scotland, UK\\
$^8$ Department of Computer Science, Aalto University, P. O. Box 15400, Espoo, FI-00076, Uusimaa, Finland \\
$^{9}$ Department of Physics, University of Helsinki, P. O. Box 64, Helsinki, FI-00014, Uusimaa, Finland
}

\date{Accepted XXX. Received YYY; in original form ZZZ}

\pubyear{xxxx}

\begin{document}
\label{firstpage}
\pagerange{\pageref{firstpage}--\pageref{lastpage}}
\maketitle

\begin{abstract}
Galaxy groups contain the majority of bound mass with a significant portion of baryons due to the combination of halo mass and abundance \citep{Cui2024}. Hence they serve as a crucial missing piece in the puzzle of galaxy formation and the evolution of large-scale structures in the Universe. In observations, mass-complete group catalogues are normally derived from galaxy redshift surveys detected through various three-dimensional group-finding algorithms. Confirming the reality of such groups, particularly in the X-rays, is critical for ensuring robust studies of galaxy evolution in these environments. Recent works have reported numerous optical groups that are X-ray undetected \citep[see, e.g.,][]{Popesso2024}, sparking debates regarding the reasons for the unexpectedly low hot gas fraction in galaxy groups. To address this issue, we utilise zoomed-in simulations of galaxy groups from the novel $\hyenas$ project to explore the range of hot gas fractions within galaxy groups and investigate the intrinsic factors behind the observed variability in X-ray emission. We find that the halo formation time can play a critical role -- we see that groups in halos that formed earlier exhibit up to an order of magnitude brighter X-ray luminosities compared to those formed later. This suggests that undetected X-ray groups are preferentially late-formed halos and highlights the connection between gas fraction and halo formation time in galaxy groups. Accounting for these biases in galaxy group identification is essential for advancing our understanding of galaxy formation and achieving precision in cosmological studies.
\end{abstract}

\begin{keywords}
keyword1 -- keyword2 -- keyword3
\end{keywords}

%%%%%%%%%%%%%%%%%%%%%%%%%%%%%%%%%%%%%%%%%%%%%%%%%%

%%%%%%%%%%%%%%%%% BODY OF PAPER %%%%%%%%%%%%%%%%%%

\section{Introduction}\label{sec1}

Galaxy groups, typically with a halo mass in the range of $\sim 10^{13} - 10^{14} M_{\odot}$ \citep{Liang2016}, contain the majority of galaxies in the Universe and serve as the primary environment for key galaxy evolution processes such as galaxy transformation driven by the interplay between galaxies and their surrounding gaseous halos \citep{OSullivan2017}. Galaxy groups offer valuable insights into fundamental physical processes such as galaxy quenching via active galactic nuclei (AGN) feedback \citep[see, e.g.,][]{Bahar2024,Yang2024,Eckert2024}, heating and cooling of the intragroup medium \citep{oppenheimer_simulating_2021}, the diverse kinematical and morphological properties of their central galaxies \citep[e.g.,][]{Loubser2018,Jung2022}, and the departure from self-similarity observed in clusters \citep[see, e.g.,][]{Yang2022}. Furthermore, galaxy groups offer constraints on galaxy formation, cosmological parameters, black hole--galaxy co-evolution, and environmental transformation \citep[see, e.g.,][for recent reviews]{lovisari_scaling_2021,eckert_feedback_2021,oppenheimer_simulating_2021}.

Despite their importance, galaxy groups have not received as much attention as galaxy clusters, mainly due to the difficulty in their detection. They are faint in the X-rays, with the temperature of the diffuse gas typically ranging from approximately 0.3 keV to 2 keV \citep[e.g.,][]{mulchaey_x-ray_2000,Liang2016}, resulting in X-ray luminosities typically ranging from $\sim 10^{40} - 10^{43}$~erg/s which is several orders of magnitude lower than clusters  \citep[][and references therein]{lovisari_scaling_2021}. Additionally, they contain far fewer member galaxies than clusters, typically ranging from a few to several dozen \citep{George2011}, which makes them more challenging to identify robustly via group-finding algorithms.  This can further lead to false identifications due to chance projections \citep[see, e.g.,][]{Pearson2017,Li2022b}. Detecting hot gas in low-mass galaxy groups (i.e., $M_h \in 10^{12.5-13.5} M_{\odot}$) is often regarded as the gold-standard for validation, but their X-ray faintness limits the number of verified low-mass groups and biases them towards lower redshifts ($z<0.4$) \citep[see, e.g.,][]{OSullivan2017,Gozaliasl2019}. Although optical and X-ray surveys continue to improve, it is critical to understand any biases introduced by group selection in order to interpret observations properly.

As an example, the scatter in their properties can introduce Malmquist biases in the inferred physical characteristics \citep{gozaliasl2020}. \cite{Damsted2023} noted a significant increase in the scatter of $L_X$ compared to other mass proxies below a redshift of 0.15, primarily in low-mass clusters, which hampers the effectiveness of X-ray observations in providing a comprehensive understanding of these groups. Recent studies \cite{Khalil2024} corroborated these findings using the AXES-2MRS galaxy groups, which combined data from the ROSAT All-Sky Survey (RASS) with the Two Micron Redshift Survey (2MRS) Bayesian Group Catalogue. They further suggested that both feedback mechanisms and halo concentration are the reasons for the substantial scatter in the properties of X-ray groups, emphasising that the scatter of scaling relations offers valuable insights into the underlying physics of galaxy groups.

The large variations in the X-ray brightness among galaxy groups can result in legitimately significant groups remaining undetected in X-ray surveys. It has been proposed that only galaxy groups with a central elliptical galaxy tend to exhibit diffused X-ray emission \citep[e.g.,][and references therein]{Mulchaey2003}, a phenomenon contingent upon the detection limits of X-ray telescopes. Utilising the Chandra X-ray Observatory, \cite{Pearson2017} investigated 10 relaxed galaxy groups carefully selected from the GAMA optical galaxy catalogue to mitigate spurious and projection effects. They observed that nine out of ten groups were underluminous in X-rays by a mean factor of approximately 4 compared to typical X-ray-selected samples. Hence, the converse practice of identifying X-ray samples and then seeking their counterparts in optically selected group catalogues may also introduce biases. Recent work by \cite{Damsted2024} expanded the findings of  \cite{Manolopoulou2021} from galaxy clusters to galaxy groups -- galaxy clusters/groups in overdense environments tend to have higher X-ray luminosities, which they hypothesised is driven by halo assembly bias. Another recent work by \cite{Popesso2024} directly combined data from eROSITA with the updated GAMA catalogue, revealing that 157 out of 189 systems with $M_{200} \geq 10^{13} M_{\odot}$ and $z < 0.2$ remained undetected in X-rays. Hence, there are significant biases introduced either when selecting groups in the optical or the X-ray \citep[see recent findings, e.g.,][]{OSullivan2017,Popesso2024}. Quantifying these biases and understanding their physical origin is essential for groups to be leveraged for galaxy formation and cosmological studies.

In this work, we examine the physical origin of the scatter in properties of galaxy groups using the \hyenas\ suite of group-scale zoom simulations (see \S\ref{sec:hyenas} for details).  \hyenas\ is a new suite that re-simulates 120 group-size halos drawn from a large-volume cosmological simulation employing the successful \simba\ galaxy formation model \citep{Dave2019}.  Its novelty lies in its selection, which is based on bins in both halo mass and halo formation time.  The latter is often implicated as a key driver in the scatter in galaxy group properties \cite[see, e.g.][]{Cui2021}.  Here we investigate what implications the variations in group halo formation times can have on their detectability in X-ray and optical surveys, and thereby quantify associated selection biases.

\section{The $\hyenas$ project} \label{sec:hyenas}

The $\hyenas$ project is a branch of the $\simba$ with its focus on galaxy groups using the zoom re-simulation technique.  While there have been many cluster-scale zoom projects, for example, the 300 project \citep[][]{Cui2018}, group-scale zooms are less common.  One reason is that large-volume cosmological simulations already contain many groups.  However, when selected carefully, zooms can sample outliers in the distribution that are not well represented in a random sample.  Also, zooms offer the opportunity to achieve higher resolution at a modest computational cost, enabling resolution convergence studies, though, in this introductory work, we do not employ that aspect of $\hyenas$.  

Besides selecting zoom halos in the group mass regime, $\hyenas$ further selects objects with a wide range in halo formation times.  This is motivated by \citet{Cui2021} who argued that halo formation time is the key determinant of the scatter in the stellar-to-halo mass relation, as well as the cold vs. hot gas content of halos.  This will presumably also impact the X-ray properties of these systems, which is relevant for this work.  Next, we describe the sample selection and X-ray analysis of the $\hyenas$ zoom suite.

\subsection{The $\hyenas$ sample and IC generation}\label{subsec:IC}

\begin{table}
	\centering
	\caption{The cosmology parameters used in the \simba simulations.}
	\label{tab:cosmology}
	\begin{tabular}{cccccc} % four columns, alignment for each
		\hline
		H0 [$km/s/Mpc$] & $\Omega_\Lambda$ & $\Omega_m$ & $\Omega_b$ & $\sigma_8$ & $n_s$\\
		\hline
		68 & 0.7 & 0.3 & 0.048 & 0.82 & 0.97\\
		% 2 & 4 & 6 & 8\\
		% 3 & 5 & 7 & 9\\
		\hline
	\end{tabular}
\end{table}

\begin{figure*}
	\includegraphics[width=\linewidth]{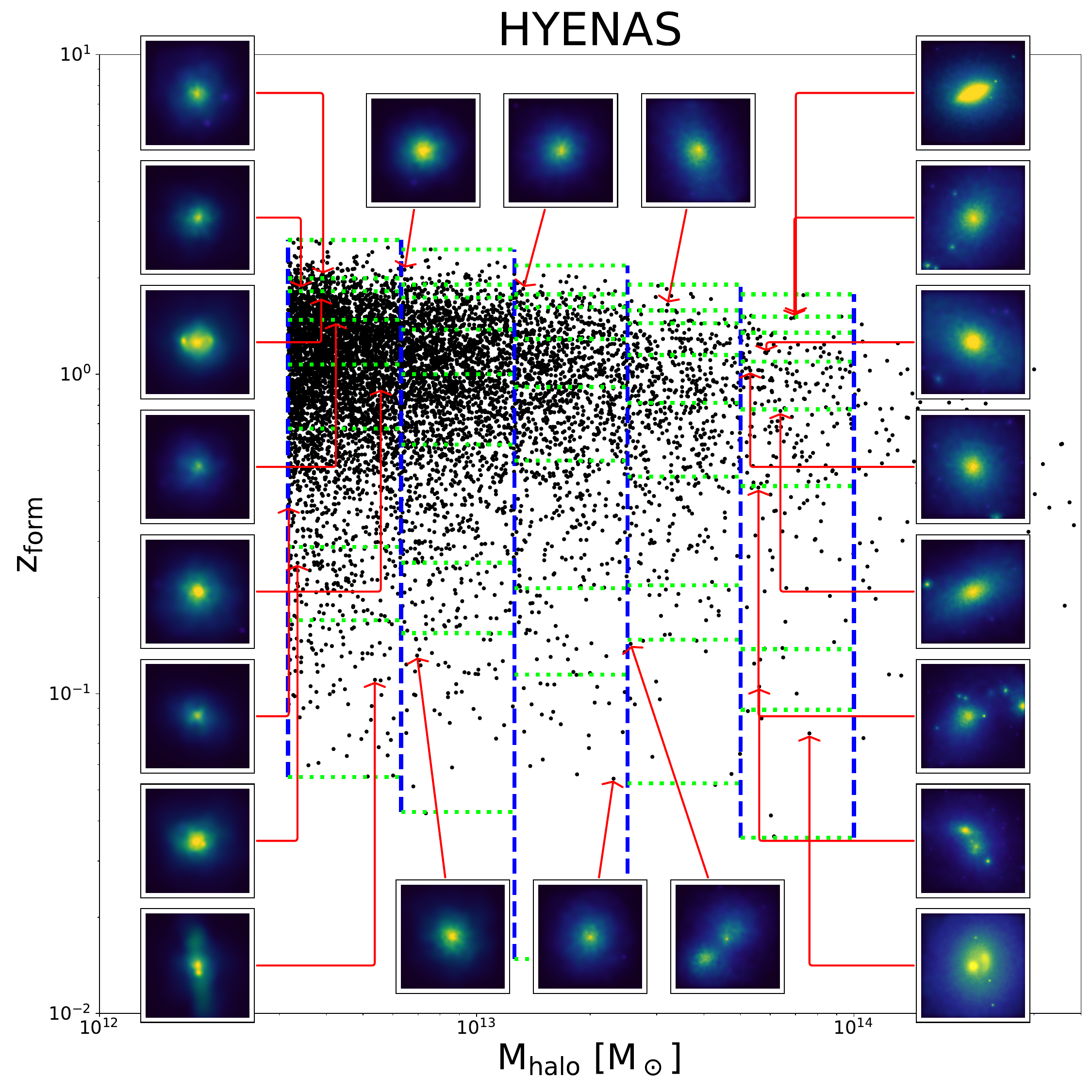}
    \caption{The halo mass -- formation time relation from the parent dark-matter-only simulation with illustrations of the selected \hyenas\ sample. The vertical blue dashed lines mark the 5 halo mass bins, while the horizontal green dotted lines indicate the $z_{form}$ percentile bins. The inset images are the blended gas and star distributions around the central galaxy of the selected \hyenas\ samples using the {\textsc Py-SPHViewer} package.}
    \label{fig:ills}
\end{figure*}

The simulation code and parameter choices used are identical to that in the \simba\ simulation, described in \cite{Dave2019} and many subsequent papers. For brevity, we do not repeat these here but focus on the aspects novel to the $\hyenas$ zoom suite.

To increase the sample of galaxy groups, we first run a $200 \hMpc$ dark-matter-only simulation ($8\times$ \simba's volume), with the same dark matter particle mass resolution and {\it Planck}-concordant cosmology (see \autoref{tab:cosmology}) as \simba using Gadget-4 \citep{Springel2021}.  From this we select 120 (out of $\sim$10k) halos with $M_{200c} \in 10^{12.5} - 10^{14} \Msun$ (where `200c' denotes 200 times the critical density). However, these are not selected randomly; rather, within each 0.5-dex mass bin, we select the galaxies covering a spread halo formation times in percentile bins.  The formation time is calculated as the time when half the $z=0$ halo mass has assembled within the halo's main progenitor.  Gadget-4's on-the-fly merger tree generation makes this calculation straightforward.

Figure~\ref{fig:ills} shows the sample selected in 5 halo mass bins and 8 formation time bins, with the latter chosen with percentiles bounds of 0-2-5-20-50-80-95-98-100 within each mass bin, and are marked as green dotted horizontal lines in \autoref{fig:ills}.  Inside each region in this space, we randomly select three halos to re-simulate. The central galaxy's density map of one example within each region is shown as an inset image. Note that this is only for illustration, as the image covers 4 times the galaxy's half-mass radii by blending both stars and dark matter with arbitrary normalisation using the {\textsc Py-SPHViewer}.

The large N-body volume's random initial conditions are generated using \music~\citep{Hahn2011}, which conveniently allows us to generate these zoomed-in ICs for the \hyenas\ sample using the same underlying white noise file. For each selected halo, we track all these dark matter particles within the halo at $z=0$ to their initial condition positions. To precisely identify the centre of the resimulation region in the IC, we also track these particles lying in the centre (minimum potential positions) of the halo ($5 \hkpc $) and use the mean position of these central particles within the IC as the `ref\_center' parameter for \music. Then, we calculate the distances from the IC central position to all the halo particle positions. The 2 times maximum distance rescaled to the simulation boxsize is used for the `ref\_extent' parameter for \music\ to make sure the interested central halo is out of contamination. %As mentioned before, the white noise at level 11 is the same as the dark-matter-only simulation. 

Each zoomed-in halo has its ICs generated with three different levels of resolution: Level 0 cuts out the zoomed-in region in the IC, adds gas particles, and decreases the resolution outside of the zoomed-in region, which is controlled by the \music\ code.  This results in a dark matter particle mass of $6.513 \times 10^7 \hMsun$ and gas element mass of $1.241 \times 10^7 \hMsun$, and a minimum gravitational softening of 0.5 $\hkpc$.  Level 1 increases the zoomed-in region's resolution by 1 level higher with a new white noise at this resolution level. The white noise is consistent as Level0 for these low-resolution levels decreased outside the zoomed-in region.  The Level 1 suite has $8\times$ lower particle masses and $2\times$ lower minimum softening.  Level 2 has a resolution that is higher than Level 1, with the outside low-resolution region being consistent with Level 1. Thus, each of the 120 selected \hyenas\ halo has 3 ICs with different resolutions. Though all the 120 \hyenas\ halos have both dark-matter-only and hydro ICs, we only run the hydrodynamic for the 40 selected `elite' halo for Level 1 and Level 2, which is still not fully finished due to their very high computation cost. The `elite' sample is the one out of the three random samples with the smallest number of high-resolution particles.

%\subsection{The $\simba$ model and the zoomed-in runs}

% Need to copy-and-past and describe the simba model here, though nothing has changed from the 100 Mpc/h run. Need to know these scaling changes, such as softening length, seeding mass for BH...

\subsection{The HYENAS catalogue and analysis}

We output 151 snapshots for each simulation run from $z=20$ to $z=0$.
Besides the on-fly FoF catalogue from \GIZMO\ and the \caesar\ catalogue based on it, we also run the AHF halo finder \citep{Knollmann2009} to identify the halos within $R_{200c}$ and produce an AHF-Caesar catalogue following \cite{Chen2024}. 

Although there are other (uncontaminated) halos inside the high-resolution zoomed-in regions, we only focus on these originally selected \hyenas\ halos in this paper. It is also recommended for all papers using the \hyenas\ data. As such, we need to identify these selected halos properly. Unfortunately, the particle IDs of these halos chosen from the parent dark-matter-only simulation are scrambled within these zoom ICs generated by \music. 
%Actually, this is true for other codes as well, it is impossible to retain this information when different levels of resolutions are used. 
Thus, we adopted another method to cross-match these halos based on the particle IDs of these zoom ICs. Since we know the exact halo centre position in these ICs, we use the IDs of dark matter particles within 300 $\hkpc$ (more-or-less corresponding to the $5 \hkpc$ radius of halos central regions at $z=0$) and track them down to $z=0$ to get their median positions. This position's distances to the $z=0$ halo centres are used to select the closest halo. If the halo mass difference from the original N-body one is less than 1.5 times, the halo is matched. If not, we choose the halo within the distance of R200c with the closest mass as the matched halo. Only seldom does this happen for these late-formed halos due to the slight evolution difference between dark matter and hydro runs. After finding the matched halos for both FoF and AHF catalogues, we compared the matched halo masses, which basically agree with previous findings \citep[for example][minor effects due to the baryons]{Cui2012,Cui2014}. We further calculate the formation redshifts using both FoF and AHF halo catalogues, which are compared to the original halo formation redshift. There is a slightly larger scatter compared to the halo mass differences, primarily coming from the $z_{form} \lesssim 0.3$ sample.

\subsection{X-ray luminosities}

X-ray properties are calculated using \moxha package \citep{Jennings2023}, which combines the yT-based PyXSIM \citep{BiffiDolag2012, BiffiDolag2013, ZuHoneBiffi2014} and Caesar \citep{2011ApJS..192....9T} software packages with the XSPEC spectral fitting package to provide an end-to-end pipeline for creating mock X-ray photon maps and analysing them to obtain mock observations such as X-ray luminosities, temperatures, and metallicities. First, we make a cut on the cold gas to remove ISM particles, which are artificially pressurised to resolve the Jeans mass \citep{Dave2019}. We use a cut such that only gas particles with a density $\rho < 0.1$ m$_\text{p}$ cm$^{-3}$ and temperature $ T > 2 \times 10^{5}$ K are included. Furthermore, we remove all wind particles and all particles with a non-zero star formation rate. We then use PyXSIM to generate X-ray emission fields in the source band of $0.5-2.0$ keV, using a CIE APEC model \citep{SmithBrickhouse2001} and using the \simba-tracked particle mass fractions for \textit{He, C, N, O, Ne, Mg, Si, S, Ca, Fe} scaled to the \citet{AndersGrevesse1989} solar abundances table. The other elements are fixed at their solar abundance ratios. We finally sum the luminosity of the hot gas particles within a radius of $R_{500}$ to give our value of $L_{X, 0.5-2.0}$. For more details, we refer to Jennings et al. 2024 in prep.
% \textcolor{red}{Weiguang, you are using the R500 value right?. Yes, that is right.}

\section{Results}\label{sec:results}
\subsection{Hot gas fraction}
\begin{figure*}
    \centering
    \includegraphics[width=\linewidth]{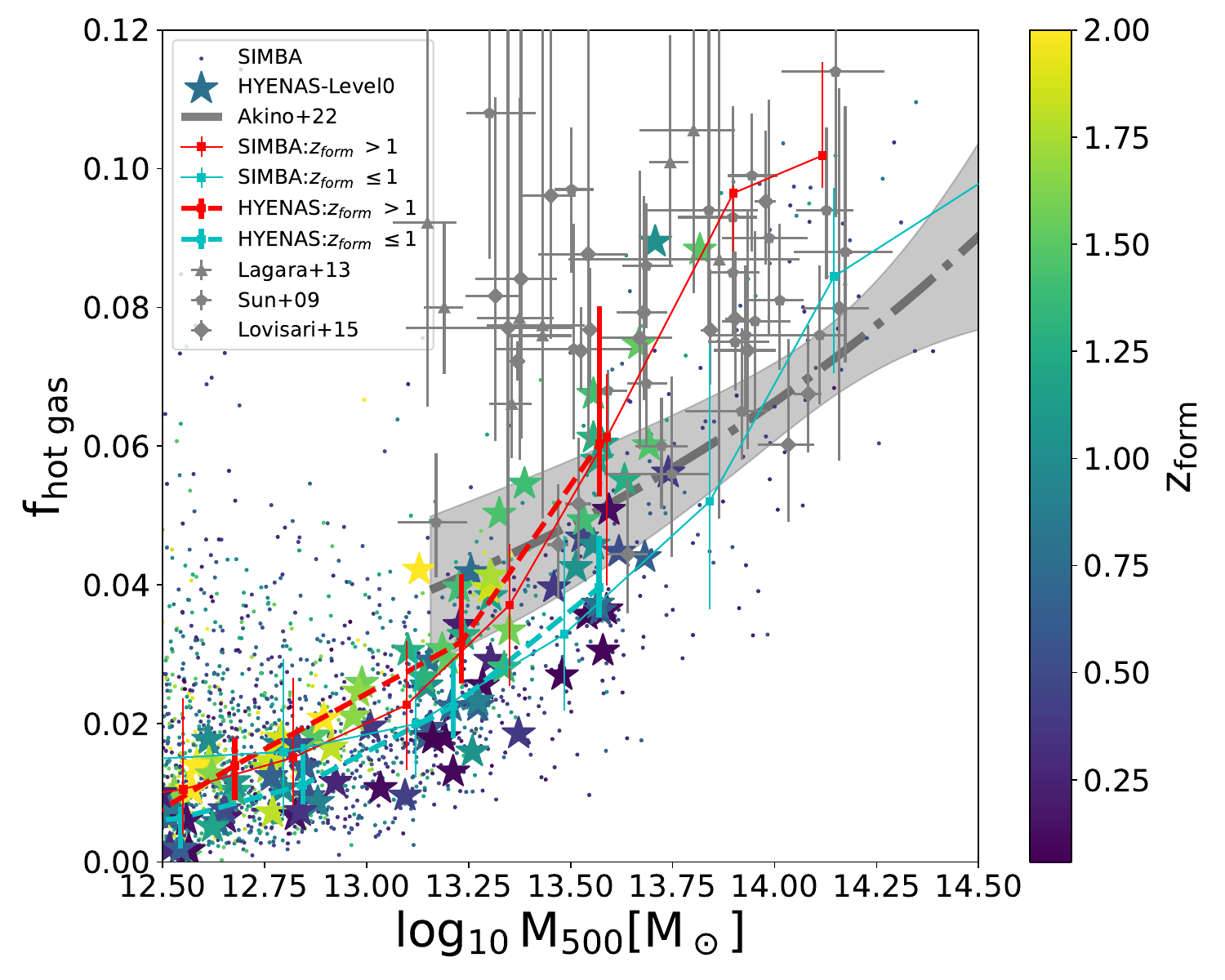}
    \caption{The hot gas fraction as a function of halo mass with colours coding to the halo formation time. The \simba\ simulation results are shown as small dots with the median values in solid thin red and cyan lines for early-formed ($z_{form}>1$) and late-formed groups ($z_{form} \leq 1$), respectively. The error bars present the $16^{th}-84^{th}$ percentile.  While the \hyenas\ sample at Level0 is indicated by stars with the medians in thick dashed lines. The observational results from \citealt{Sun2009,Lagana2013,Lovisari2015,Akino2022} are shown in grey symbols with error bars and a dotted-dashed line with a shadow region as indicated in the legend. }
    \label{fig:fgas}
\end{figure*}

Since the X-ray is coming from hot gas, we first investigate the hot gas mass fraction in both the galaxy groups from both \hyenas\ (large stars) and \simba\ (small points) simulations and compare them to recent observations from \cite{Sun2009,Lagana2013,Akino2022} in \autoref{fig:fgas}. First, in agreement with observation, there is a decreasing trend for the $f_{hot\ gas}$ with decreased halo mass. The simulations agree better to the observation data points from \cite{Sun2009,Lagana2013} \citep[as well as][]{Liang2016} than those of \cite{Akino2022} at $M_{500} \gtrsim 10^{13.5} \Msun$. Note that the gas fraction in the massive halo seems to agree better with \cite{Akino2022} as shown in \cite{Cui2018}, which could be due to the different AGN feedback strengths. While in the intermediate halo mass range, the simulation data tends to agree with \cite{Akino2022} instead of others \citep[see also][]{Robson2020}. The disagreements between these observations' results at the galaxy group scale have been discussed in \cite{eckert_feedback_2021}. Here we would like to add an additional potential cause: \cite{Akino2022} and \cite{Eckert2016} (which also has a lower gas fraction compared to the others) are both based on the XXL survey \citep{Pierre2016}, which is a volume complete survey, while the others mostly preferentially selected the X-ray bright objects. At lower halo mass with $M_{500} \lesssim 10^{13}$, this trend flattens. 

Secondly, there is a large scatter of the data points in \autoref{fig:fgas}. It is worth noting that the \hyenas\ sample generally covers the distribution of \simba\ data points well at $M_{500} \gtrsim 10^{13} \Msun$. These outliers from \simba\ with $f_{hot\ gas} \gtrsim 0.04$ at $M_{500} \lesssim 10^{13} \Msun$ could be due to the fact that they are close to a massive cluster, as discussed in \cite{Cui2022}; in contrast, \hyenas\ groups are selected to be isolated halos by design. Nevertheless, the agreement of overall distributions between \simba\ and \hyenas\ suggests that our $\hyenas$ selection criteria are unbiased with respect to halo gas fractions. 

The colours of $\hyenas$ points indicate the halo formation redshift as shown by the colourbar in \autoref{fig:fgas}.  This shows an interesting trend -- the gas fraction at a given halo mass correlates with its halo formation time. Late-formed halos tend to have a smaller gas fraction. To statistically show this trend, we separate the data points into $z_{form} > 1$ (red lines) and $z_{form}\leq 1$ (cyan lines). There is quite good agreement between the thick (\hyenas) and thin (\simba) lines, and both have a clear separation between early-formed and late-formed halos.

The halo formation time dependence, we speculate, owes to accumulated heating processes from both AGN feedback and shock heating from structure formation. For early-formed halos, it will not only have an early accretion of more cold gas at very high redshift \citep[][]{Cui2021}, experience shock heating earlier and longer but also form its central galaxy earlier with a massive black hole according to the $M_*- M_\cdot$ relation, which leads to an earlier jet mode feedback in the \simba\ model. 
% This strong feedback blows gas particles with a speed of $7000 km/s$, which not only drives the quenching of the central galaxy but also heats the CGM or retains its hot temperature by counteracting cooling~(Jennings et al. 2024 in prep.). 

% One would argue that the longer AGN feedback will reduce the hot gas fraction by blowing them outside of $R_{500}$ instead of increasing it. However, we would like to add here that the central black hole mass at this halo mass range is still quite low \citep{Dave2019} to allow a longer time of jet mode AGN feedback in \simba. As such, some central galaxies are not quenched yet. This leaves the key heating process is unclear 

At lower halo mass $M_{500}\lesssim10^{13}\Msun$, the late-formed halos tend to have a higher gas fraction than early-formed halos, i.e., a reversed trend, albeit a large error bar for the \simba\ simulation. In comparison, the cross point is at a relatively lower halo mass, $M_{500} \approx 10^{12.5} \Msun$, for \hyenas, which could be due to the limited number of objects. At this lower halo mass, the jet mode AGN feedback will not turn on because the BH mass \citep[see][for the halo mass -- BH mass relation]{Cui2022} is lower than the threshold set in \simba\, which is around $10^8 \Msun$.
% (WEIGUANG: Not sure about this, I think all these halos are well above that threshold) (Reply: I don't mean the resolution here, it is the MBH threshold for jet feedback is allowed, I think it is around 1e8 Msun in simba, which corresponds to the halo mass around 1e12.5 Msun). 
Thus, the heating process should be dominated by the supernovae feedback, which can be more dominant in these late-formed halos with higher star formation. We will return to these explanations in the discussion part in detail.

\subsection{X-ray luminosities}
\begin{figure*}
    \centering
    \includegraphics[width=\linewidth]{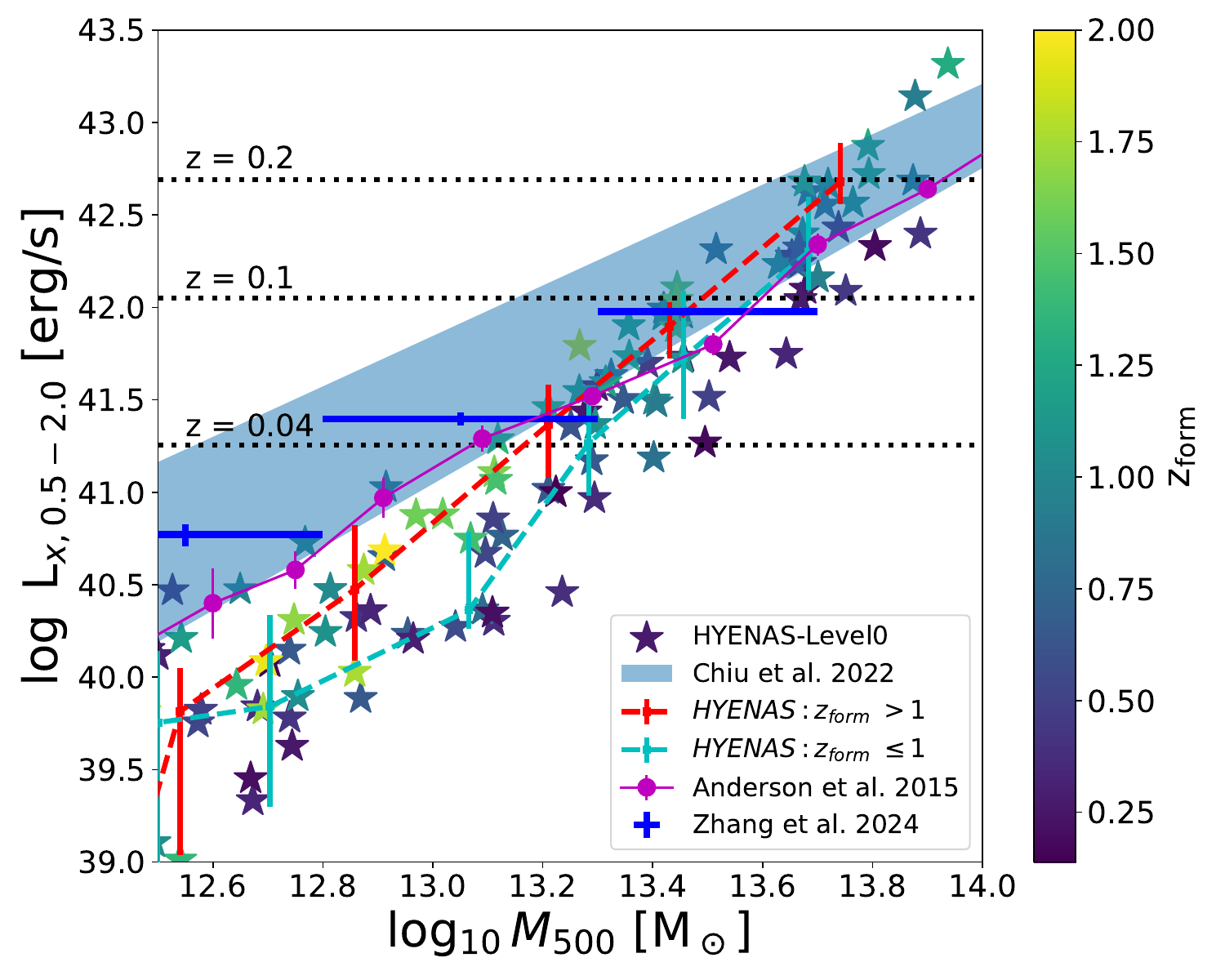}
    \caption{Similar to \autoref{fig:fgas}, but for the groups X-ray luminosity. Magenta lines and symbols show the result from \citealt{Anderson2015} with the X-ray luminosity estimated at the same energy band -- 0.5 - 2.0 keV. Blue error bars are the results from \citealt{Zhang2024} using the latest eROSITA survey catalogue. We remind here again that the errorbars are different in simulation ($16^{th}-84^{th}$ percentiles) from observations.}
    \label{fig:x-ray}
\end{figure*}

%Previous hot gas fraction is only a theoretical and rough estimate to understand how the X-ray luminosity can be affected. Thanks to \moxha\citep{Jennings2023}, we are able to mimic the whole observation process and make direct, consistent investigations on why some groups are X-ray faint or even undetected. 

In \autoref{fig:x-ray}, we show the \hyenas\ groups' X-ray luminosity within the soft band, [0.5, 2] keV, which has been adopted by many surveys for detecting X-ray galaxy groups. Similar to the hot gas fraction shown in the previous section, the late-formed halos have a lower X-ray luminosity than early-formed ones. The differences between the two families at the same halo mass can be an order of magnitude. Separating the sample by their $z_{form}$ with the arbitrarily selected threshold of $z=1$, we observe the same trend as shown in \autoref{fig:fgas} with slightly larger separation at $M_{500}\approx 10^{13}$, then a reversed trend at lower halo mass $M_{500} \approx 10^{12.5}\Msun$. As indicated by \cite{Bulbul2024}, the detection limit of eRosita at $z \approx 0.2$ (the upper redshift limit in \citep{Popesso2024}) is around $5\times 10^{42}$ ergs/s, which is indicated by the top horizontal dotted line in \autoref{fig:x-ray}. If we naively use that as the detection limit of eROSITA at $z =0.1$, it is clear that some of our simulated halos, even with $M_{500} \sim 5\times 10^{13} \Msun$ can not be seen. 
% This explains the missing X-ray emission of the X-GAP galaxy group (Eckert et al. in prep.).

In comparison, we include the observation results from \cite{Anderson2015}, which are based on stacking X-ray emission from the ROSAT All-Sky Survey around the local brightest galaxies. The halo mass in \cite{Anderson2015} is computed using the simulated catalogue of these local brightest galaxies. 
It is clear that the magenta line is lying in between the \hyenas\ samples at $M_{500} \gtrsim 10^{13.2} \Msun$, roughly crossing the detection limit, and consistently lying on the top boundary of the \hyenas\ low mass sample. Similarly, \cite{Zhang2024} stack the X-ray luminosities of the CENhalo sample, which is binned in halo mass $M_{200m}$ based on the group finder algorithm \citep{Tinker2021}. We used the corresponding $M_{500c}$ in \autoref{fig:x-ray}, which is derived considering the concentration model from \cite{Ishiyama2021}. Note that the uncertainty on $L_X$ is estimated from the quadratic sum of the Poisson error, which is not the same as what we are showing in \hyenas. Instead of stacking the galaxy groups, \cite{Chiu2022} removed these contaminated systems due to a random superposition. With these remaining 434 groups and clusters, which are cross-confirmed via their weak lensing masses from the HSC survey, they did an MCMC fitting and yielded a similar result to \cite{Zhang2024} as shown in \autoref{fig:x-ray}. In conclusion, these observation results in the galaxy group scale basically lie on the upper end of the \hyenas\ result, which could be explained if the observations are missing these X-ray faint groups.

\begin{table}
\caption{Predicted X-ray detection limits and fractions}\label{tab1}%
\begin{threeparttable}
\begin{tabular}{@{}lllll@{}}
\toprule
redshift & min mass\tnote{1} & max mass\tnote{2}  & mean mass\tnote{3} & fraction\tnote{4}\\
z & $\log M_{500}$ & $\log M_{500}$ & $\log M_{500}$ & 100\% \\
\midrule
$0.2$ & 13.676   & 13.874  & 13.775 & 15.4  \\
$0.1$ & 13.444   & 13.751  & 13.597 & 71.4  \\
$0.04$ & 13.118  & 13.495  & 13.307 & 70.0 \\
\bottomrule
\end{tabular}
  \begin{tablenotes}
  \item[1] The minimum halo mass in $\Msun$ above the X-ray luminosity limit at a given redshift, corresponds to 0 per cent detection if the halo mass is smaller than this one.
\item[2] The maximum halo mass above the X-ray luminosity limit at a given redshift, corresponds to 100 per cent detection if the halo mass is larger than this one.
\item[3] The mean of columns 2 and 3 for calculating the detection fraction in column 4.
\item[4] The fraction of X-ray detected groups within the halo mass bin, column 3 $\pm 0.1$.
  \end{tablenotes}
\end{threeparttable}
\end{table}

\noindent
Based on \hyenas\ data in \autoref{fig:x-ray}, we further list the roughly predicted limits for eROSITA in \autoref{tab1}. At each of the three redshifts, we first select all \hyenas\ halos within $\pm$ 10 per cent of the X-ray detection limit. The minimum and maximum halo masses within that $L_x$ limit are listed in the second and third columns of \autoref{tab1}, respectively. Using the mean (column 4) of columns 2 and 3, we select all halos within a halo mass bin of $\pm 0.1$, then give the detection fraction in the fifth column of \autoref{tab1} as $N_{L_x > L_{x,\ lim}}/N_{total\ in\ bin}$. We note here that the \hyenas\ sample is not a mass-complete one. Therefore, these limits and fractions only serve as a rough prediction. The clear drop of the fraction at $z=0.2$ presents a good agreement to \cite{Popesso2024}. However, we will need a larger sample to confirm this.
% \subsection{Connection to central galaxy properties}

% Need to study the CGG SFR, etc.

\section{Discussion and conclusion}\label{sec:discussion}

As we have presented before, it is clear that the X-ray-detected galaxy groups are biased toward these gas-rich ones, which are closely linked to their early halo formation time. Although it is easy to understand that -- early halo formation will bring more cold gas at high redshift as shown in \cite{Cui2021}, those gases will be heated up by either shock heating in structure formation or feedback in the process of galaxy formation -- it is unclear how this fits into the picture of general expectations from galaxy formation. We further break this down into three aspects:
\begin{itemize}
    \item[$\bullet$] {\bf [Connection to the central galaxy]}
    Observations have suggested that early-type elliptical galaxies in galaxy groups tend to be associated with diffuse X-ray emission, while late-type disk galaxies do not \citep{Mulchaey2003}. This is especially interesting because \cite{Cui2021} studied the connection between the central galaxy stellar mass and the halo mass and revealed the scatter in that relation is intrinsically driven by halo formation time. In their Supplementary Figure 2, it is clear that early-formed halos tend to host massive quenched galaxies at the same halo mass when $M_{halo}>10^{13} \Msun$, reversed from the low mass halos in which the red/quenched galaxies with lower stellar mass tend to live in late-formed halos. This is further confirmed by the galaxy age (see \autoref{fig:fgas_bgg_age}) if we note that the early-type galaxies are older. We understand that this issue is still in debate; see \cite{Scholz-Diaz2024} for the most recent discussions on that. We argue here that our results (including \cite{Cui2021}) are consistent with their claim that higher stellar mass galaxies at a given halo mass have characteristics of old, red, and passive systems {\it at halo mass larger than $10^{13}\Msun$}. While there is less data in \cite{Scholz-Diaz2024} at low halo/stellar mass range to make a solid conclusion, and their total mass is only calculated within 3$R_e$. 

    Due to being driven by halo formation time in both relations, we expect a positive connection between the gas fraction and central galaxy stellar mass at group halo mass scales -- more massive galaxies tend to be surrounded by more hot gas. This is proved in \autoref{fig:fgas_fs}. As suggested by \cite{Correa2020}, disc galaxies are less massive than elliptical galaxies in same-mass haloes when the halo mass is larger than $10^{13}\Msun$, which confirms the previous suggestion that elliptical galaxies tend to associate with X-ray emissions. 

    We further find a positive correlation between the central galaxy mass-weighted age and halo formation time, as shown in \autoref{fig:fgas_bgg_age}. This again confirms our results at $M_{500}\gtrsim 10^{13} \Msun$ are consistent with \cite{Scholz-Diaz2024} and lead to a positive correlation with the scatter in central galaxy stellar mass.  This is contrary to the findings of \cite{Kulier2019}, which could be because they used all galaxies within the EAGLE simulation, so low-mass halos dominate the sample. At low halo masses $M_{halo} \lesssim 10^{13} \Msun$, such an anti-correlation is also found in \cite{Cui2021}. We also note that a crossing point is shown in \autoref{fig:fgas}, which should be consistent with the reversed trend at low halo mass, though at slightly different halo masses when comparing \simba\ and \hyenas.

    \item[$\bullet$]  {\bf [The abundance of cold gas]} 
    In previous figures, we only focus on the hot gas mass fraction; it is unclear how the cold gas abundance will contribute to the full picture, i.e. whether the low hot gas fraction is due to a high cold gas fraction or not. This is because galaxy groups, unlike galaxy clusters, tend to host a noticeable fraction of gas mass in cold as well. Investigating that will help us to form a full picture of how they are formed. As expected, the cold gas gradually contributes more to the total gas mass with the halo mass reduced after $M_{500} \lesssim 10^{13.4} \Msun$, see \autoref{fig:fhot}. It is further interesting to see that there is more cold gas in late-formed halos than in early-formed ones in that figure, which we will discuss the reasons for in the following section. That reveals that the history of halo formation also affects the history of gas thermalization.

    \item[$\bullet$]  {\bf [Connection to central BH]} 
    Although there is more gas in early-formed halos, the gas must be hot to be seen in the X-ray band. Therefore, the heating processes are key to understanding why there is more hot gas in these early-formed halos than in late-formed ones. As shown in the previous section, the early-formed halos not only have more gas but also more hot gas than these late-formed halos (see \autoref{fig:fhot}). Thus, early-formed halos should have more heating sources/energies than late-formed ones. One possible reason is shock heating, which should happen earlier in early-formed halos, yielding a hot gas fraction. The other reason is AGN feedback. For example, \cite{Liang2016} suggested that the winds ejected from the group galaxies interact with and heat the hot halo gas, which not only reduces the rate at which the halo gas cools and accumulates in the group's central galaxies but also causes its distribution to remain more extended. More importantly, we found that the massive galaxies tend to host a more massive BH at the same halo mass \citep[see][at the more massive halo mass end; see also Ma et al. in prep.]{Davies2019,Davies2020}. This is not surprising since the early-formed halos tend to form the central galaxy earlier, and as such, the central BH mass grows faster and earlier. For the case of \simba\ model, it enters the jet mode earlier to quench the central galaxy \citep{Cui2021} with the higher hot gas mass as a by-product. This picture is supported by \autoref{fig:fgas_mbh}. We are currently working on another paper to record the heating energy from different sources to determine which is more important for gas heating in galaxy groups.
\end{itemize}

Our findings in this work are based only on the \simba\ baryon model. However, we also investigated the TNG-300 simulation, which shows the same gas fraction trend with a clear separation between early- and late-formed halos, albeit with a little systematically higher values than what is shown in \autoref{fig:fgas} \citep[see also][for the higher gas fraction in TNG than EAGLE]{Davies2020}. Recent observation work by \cite{Popesso2024}, which compared the X-ray detected and undetected groups, also suggested a similar conclusion, i.e., halo assembly bias is the cause. Furthermore, \cite{Andreon2022} showed that under X-ray luminous clusters populate the low concentration of dark matter end of the distribution for a given mass, suggesting that they are late-formed as well.  However, the halo formation redshift is very hard to measure in observations. There are ways to approximate it, such as the galaxy magnitude/stellar mass difference in fossil groups \citep[e.g.][]{Jones2003,Gozaliasl2014b} and the connection between galaxy/gas dynamical state and halo formation time \citep[e.g.][]{Mostoghiu2019}, but all have a substantial uncertainty. Though there are claims that the fossil groups show no difference to normal groups in X-ray scaling relations \citep[e.g.][]{Kundert2015, Girardi2014}, their lowest X-ray is still above $\sim 10^{42}$, which is much higher compared to the limit shown in this study. In this theoretical investigation, we don't probe into details but suggest these connections to galaxy and BH properties can be tested in observations as discussed in previous paragraphs.

Another possible explanation for these X-ray faint or undetected galaxy groups is the projection effect when they are generally identified through the galaxy catalogues \citep[see][for example]{Hernquist1995}. If two small halos are lying along the same line of sight but have a large separation, neither will have large enough hot gas to shine in X-ray, see \autoref{fig:x-ray} for how quickly the X-ray luminosity drops with halo mass. However, this projection issue may be solved by highly accurate spectroscopic redshift measurement with proper galaxy velocity distribution modelling. 
%\citep[e.g.][]{Wojtak2007}. 

Lastly, the role that baryonic physics models play in this result is not very clear, especially the AGN feedback, which may affect the X-ray luminosity. For example, \cite{KarChowdhury2022} showed that the different versions of \simba\ run turning on and off different \simba' models, especially the X-ray AGN feedback and radiative mode, result in different surface brightness profiles at different radii. However, we argue that this will only systematically shift our result, while the effect of halo formation time on the X-ray luminosity will be unchanged, which we have confirmed with the TNG-300 result. On the other hand, many studies with the EAGLE simulation show consistent predictions, as we have discussed before.

\bigskip\noindent

\section*{Acknowledgements}

WC is supported by the STFC AGP Grant ST/V000594/1, the Atracci\'{o}n de Talento Contract no. 2020-T1/TIC-19882 was granted by the Comunidad de Madrid in Spain, and the science research grants were from the China Manned Space Project. He also thanks the Ministerio de Ciencia e Innovación (Spain) for financial support under Project grant PID2021-122603NB-C21 and HORIZON EUROPE Marie Sklodowska-Curie Actions for supporting the LACEGAL-III project with grant number 101086388. 

FJ would like to acknowledge the support of the Science and Technology Facilities Council (STFC).

AB acknowledges support from the Natural Sciences and Engineering Research Council of Canada (NSERC) through its Discovery Grant program, the Infosys Foundation, via an endowed Infosys Visiting Chair Professorship at the Indian Institute of Science and the Leverhulme Trust.

These \hyenas\ simulations were performed using the DiRAC@Durham facility managed by the Institute for Computational Cosmology on behalf of the STFC DiRAC HPC Facility (www.dirac.ac.uk) with the DiRAC Project: ACSP252, titled Simba Zoom Simulations of Galaxy Groups. The equipment was funded by BEIS capital funding via STFC capital grants ST/P002293/1, ST/R002371/1 and ST/S002502/1, Durham University, and STFC operations grant ST/R000832/1. DiRAC is part of the National e-Infrastructure. The analysis reported in this paper was also partly enabled by WestGrid and the Digital Research Alliance of Canada (alliancecan.ca).

%%%%%%%%%%%%%%%%%%%%%%%%%%%%%%%%%%%%%%%%%%%%%%%%%%
\section*{Data Availability}

The data output of this paper is available on GitHub:
The \hyenas\ simulation is currently available upon request, but it will become publicly available very soon.



\bibliographystyle{mnras}
\bibliography{reference} % if your bibtex file is called example.bib

%%%%%%%%%%%%%%%%%%%%%%%%%%%%%%%%%%%%%%%%%%%%%%%%%%

%%%%%%%%%%%%%%%%% APPENDICES %%%%%%%%%%%%%%%%%%%%%

\appendix

\section{hot gas abundance}\label{secA1}
\begin{figure}
    \centering
    \includegraphics[width=\linewidth]{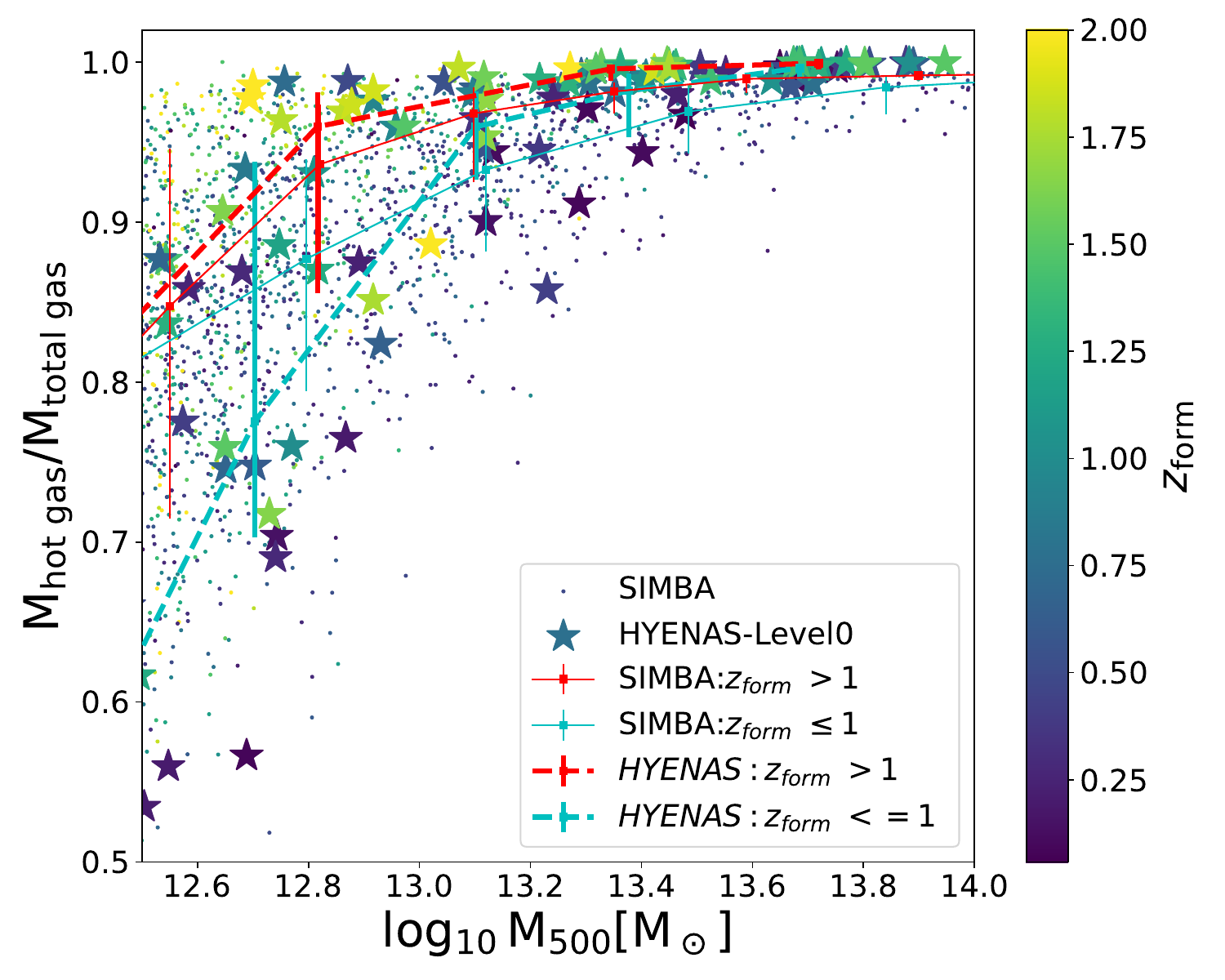}
    \caption{The hot gas mass fraction with respect to the total gas mass. The symbols and lines share the same meanings as \autoref{fig:fgas}. The hot gas dominates ($\gtrsim 0.95$) the total gas mass in halos with $M_{500} \gtrsim 10^{13.4}\Msun$. Cold gas starts to contribute more mass as the halo mass drops. It is also interesting to see that the hot gas fraction is higher in these early-formed halos than in later-formed ones at the same halo mass, which is clearer at the lower halo mass.}
    \label{fig:fhot}
\end{figure}

As the halo mass decreases, gas heating becomes weak for various reasons. As such, galaxy groups, unlike clusters, may contain a certain fraction of cold gas, which doesn't emit X-ray photons. As such, it would be interesting to understand the cold gas content in galaxy groups. In \autoref{fig:fhot}, we show the hot gas mass fraction with respect to the total gas mass. Furthermore, the simulation data is coloured and split by their halo formation time. Although hot gas still occupies the most mass in galaxy groups, the decreasing fraction is very clear along halo mass, and late-formed halos have systematically more cold gas than early-formed halos.

\section{gas fraction separated by central galaxy fraction}\label{secA2}
\begin{figure}
    \centering
    \includegraphics[width=\linewidth]{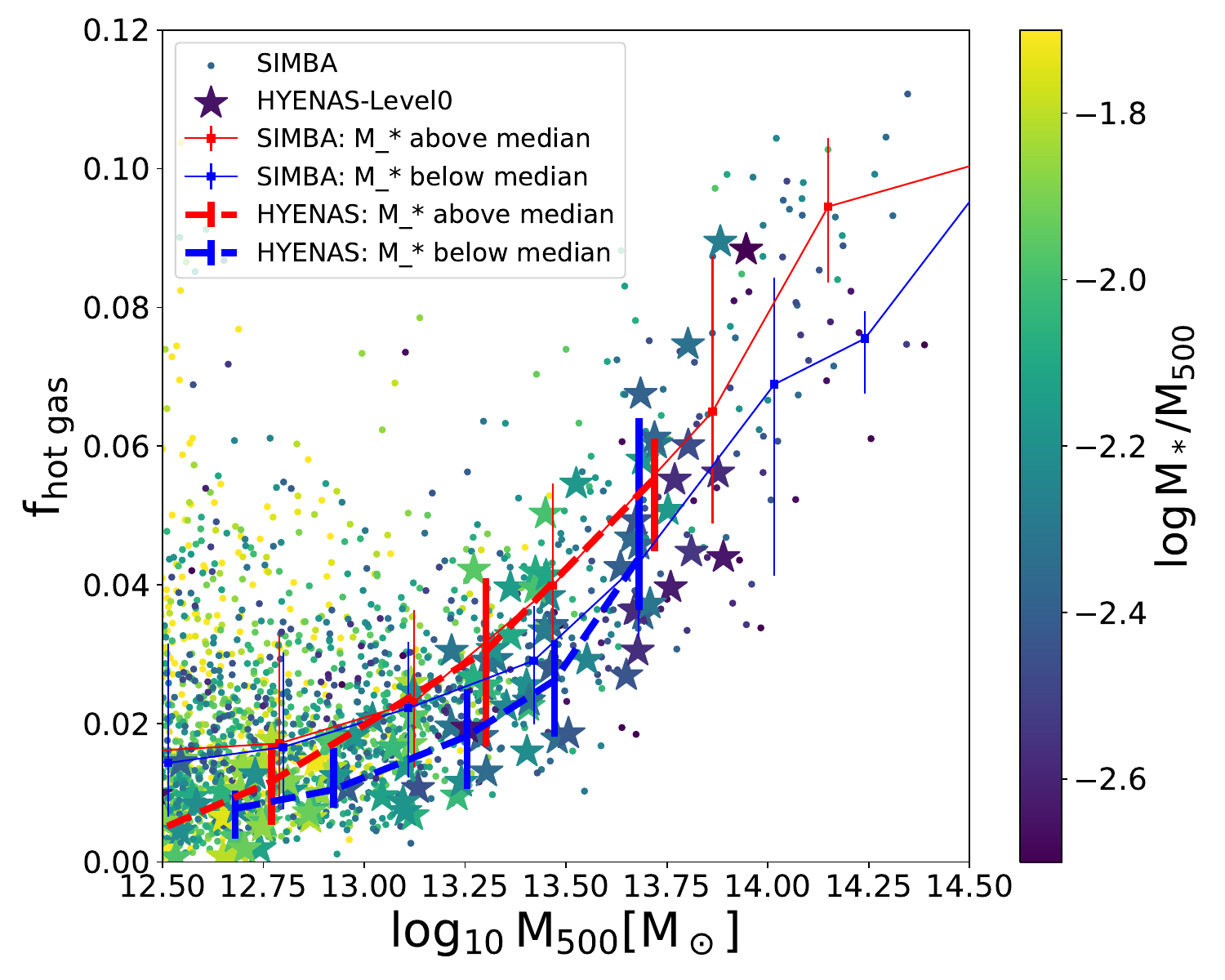}
    \caption{Similar to \autoref{fig:fgas} but color coding to the central galaxy mass fraction. Due to the stellar mass fraction depending on halo mass, we used its median value in the $M_*/M_{halo}$ - $M_{halo}$ to separate the two families. At $M_{500}$ above $10^{13}\Msun$, More massive central galaxies tend to have higher hot gas fractions than less massive galaxies at the same halo mass.}
    \label{fig:fgas_fs}
\end{figure}

Instead of halo formation time, which directly affects the central galaxy properties \citep[e.g.][]{Cui2021}, we investigate the central galaxy (or brightest group galaxy in observation, BGG) stellar mass fraction's influence on the hot gas fraction in \autoref{fig:fgas_fs}. The symbol colours of simulated objects are with respect to the BGG's stellar mass fraction -- $M_*/M_{500}$. To clearly show its effects, we first estimate the median line in the $M_{500}$ - $M_*/M_{500}$ relation. Then, separate these symbols in \autoref{fig:fgas_fs} into two groups: above the median line or below the median line in the $M_{500}$ - $M_*/M_{500}$ relation. After that, we show the median values of the two groups in red and blue, respectively. The high gas fraction is generally associated with a massive BGG.

\section{gas fraction separated by central BH mass}\label{secA3}
\begin{figure}
    \centering
    \includegraphics[width=\linewidth]{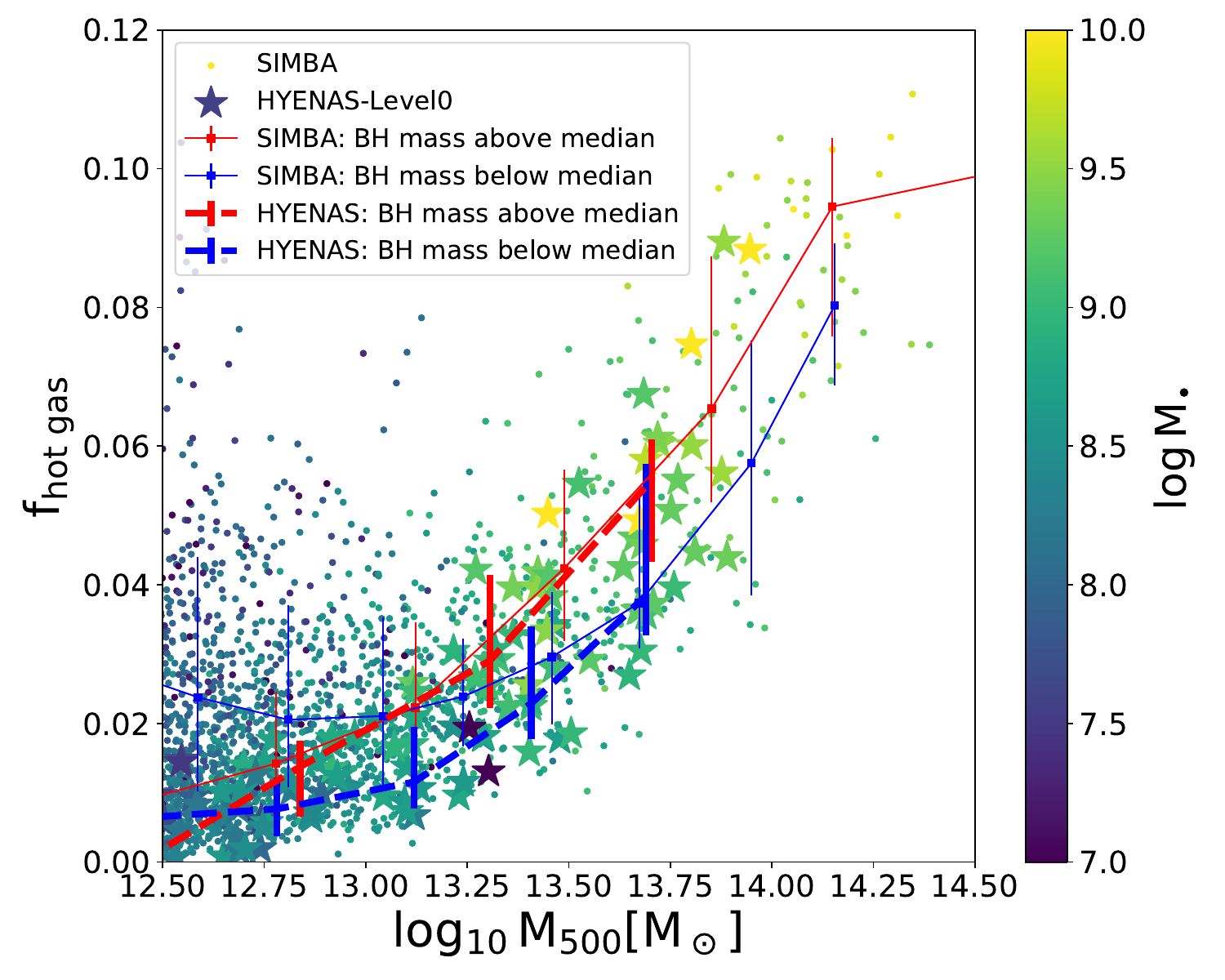}
    \caption{Similar to \autoref{fig:fgas_fs} but colour coding to the BH mass in the central galaxy. Again, we use the median line in the $M_{\bullet} - M_{halo}$ relation to split the massive and low mass BH families. A more massive central BH tends to have a higher gas fraction at the same halo mass range when $M_{500} \gtrsim 10^13 \Msun$.}
    \label{fig:fgas_mbh}
\end{figure}

Similar to \autoref{fig:fgas_fs}, we highlight the effect of black hole mass in \autoref{fig:fgas_mbh}. It is not surprising to see the halo with a higher black hole mass tends to have more hot gas. This is because we know that the black hole mass is primarily scaling with its host galaxy's stellar mass. On the other hand, this hints the AGN feedback may play a role in the higher hot gas fraction. 

\section{gas fraction separated by central galaxy age}\label{secA4}
\begin{figure}
    \centering
    \includegraphics[width=\linewidth]{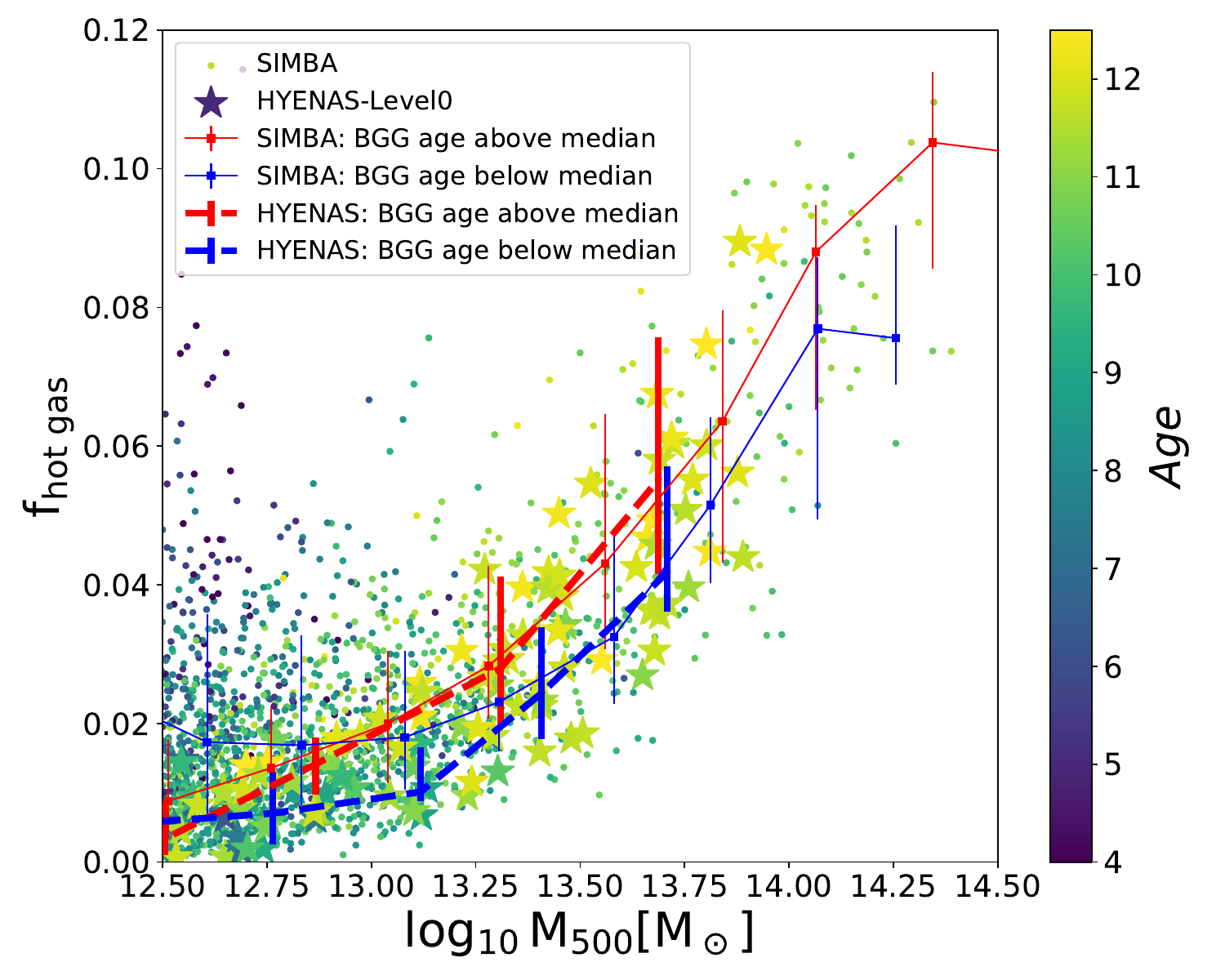}
    \caption{Similar to \autoref{fig:fgas_fs} but colour coding to the stellar age of the central galaxy. Again, we use the median line in the $age - M_{halo}$ relation to split the sample into two galaxy age families. At the same halo mass range with $M_{500} \gtrsim 10^13 \Msun$, an older galaxy tends to have a higher gas fraction.}
    \label{fig:fgas_bgg_age}
\end{figure}

Instead of BGG's stellar mass fraction, we show the connection to BGG's mass-weighted stellar age in \autoref{fig:fgas_bgg_age}. Again, this fits into the consistent picture of BGGs formed earlier with older age to have more mass in the early-formed halos. 

%%%%%%%%%%%%%%%%%%%%%%%%%%%%%%%%%%%%%%%%%%%%%%%%%%

% Don't change these lines
\bsp	% typesetting comment
\label{lastpage}
\end{document}